# Comet C/2011 J2 (LINEAR): Photometry and Stellar transit.

Oleksandra Ivanova,[a*] Serhii Borysenko,[a] Evgenij Zubko,[b,c] Zuzana Seman Krišandová,[d]

Ján Svoreň,[d] Aleksandr Baransky,[e] Maksim Gabdeev[f]

[a] Main Astronomical Observatory of National Academy of Sciences, Kyiv, Ukraine.

[*] Corresponding Author. E-mail address: sandra@mao.kiev.ua

[b] School of Natural Sciences, Far Eastern Federal University, Vladivostok, Russia

[c] Institute of Astronomical, V. N. Karazin Kharkov National University, Kharkov, Ukraine

[d] Astronomical Institute of the Slovak Academy of Sciences, SK-05960 Tatranská Lomnica, Slovak Republic

[e] Taras Shevchenko National University, Kyiv, Ukraine.

[f] Special Astrophysical Observatory, Russian Academy of Sciences, Nizhnii Arkhyz, Russia.

ABSTRACT

We present results of two-year photometric monitoring of Comet C/2011 J2 (LINEAR) that spans the time period from February 2013 through December 2014, before and after perihelion passage. The observations were conducted with broadband R filter. Analysis of observations of Comet C/2011 J2 (LINEAR) allows estimating the nucleus radius as function of geometric albedo and phase-darkening coefficient. Furthermore, our observations showed split of the comet. Fragment (B) additional to the primary C/2011 J2 nucleus was unambiguously detected; relative velocity is estimated to be ~ 0.17 ″/day. We derive the Afρ parameter and estimate the dust production rate in Comet C/2011 J2 (LINEAR) over the entire run of observations. We found a noticeable increase in Afρ parameter between September 18, 2014 and November 5, 2014, epoch when the nucleus presumably got split. On September 28, 2014, we observed a transit of the 16-magnitude star (USNO-A2 1275-18299027) near nucleus of the Comet C/2011 J2 (LINEAR). We retrieve the optical depth of the coma $\tau = 0.034 \pm 0.1$. The filling factor f that corresponds to such optical depth is in good quantitative agreement with the value that can be derived from the Afρ parameter under reasonable assumption on geometric albedo of cometary dust.

## 1. Introduction

Dust is an important species of cometary coma. The outflow of gas and dust originate from a bounded area of active regions on a surface of cometary nucleus (Belton, 2010). Such active areas were firstly discovered by VEGA-1, 2 and Giotto spacecrafts on Comet 1P/Halley (A'Hearn et al., 1986) and, later, on all the comets studied in situ (Belton, 2010), including the latest case of Comet 67P/Churyumov-Gerasimenko (Sierks et al., 2015).

Relative amount of dust and gases in a comet is characterized with the so-called dust-to-gas ratio. This parameter differs from one comet to another. Typically, it is considerably lower than 1; whereas, in some comets, the dust-to-gas ratio can be as high as ~1 (Singh et al., 1992). Note also, relative abundance of dust in comets is subject of short-term variations. Despite a relatively small amount of refractory materials in comets, dust plays an important role in general context of formation and evolution of comets. Evidently, productions of gases and dust, as well as their relative abundance, are dependent on heliocentric distance of a comet (Singh et al., 1992). However, this may also imply some changes in physical properties of cometary dust. For instance, size distribution of dust particles expelled from the nucleus is dependent on ability of subliming ices to lift them off the surface and, hence, coma is expected to be dominated by different populations of particles at different heliocentric distances of the comet. Therefore, dust in distant comets might differ from that in comets in vicinity of the Sun (e.g. Kolokolova et al., 2007; Ivanova et al., 2015).

Light scattered from a comet, in general, is contributed by elastic scattering from its dust and molecular emission from its gases. Their relative contribution into light-scattering response varies throughout coma because dust and gases follow different spatial distributions in the coma. For instance, gas molecules acquire velocities of ~$10^3$ m/s that is an order of magnitude greater compared to submicron dust particles (~$10^2$ m/s); whereas, an increase of size further decreases ejection speed of dust particle. However, different ejection velocity of dust particles and impact of the solar radiation pressure may produce noticeable spatial segregation of cometary dust particles. Furthermore, non-regular outburst activity and/or fragmentation of cometary nucleus may temporarily affect population of cometary dust particles (Hadamcik & Levasseur-Regourd, 2003; Hadamcik et al., 2007; Zubko et al., 2011). Partial fragmentation or complete disintegration of cometary nucleus may allow us to study its internal and, plausibly, pristine composition. Although numerous comets observed to date are quite stable objects, fragmentation or split of comets also is a relatively frequent phenomenon (A'Hearn et al., 1980; Sekanina et al., 1997; Bockelée-Morvan et al., 2001). Study of disintegrating comets is especially important in case of new comets because they are thought to have no experience of close approach(s) to the Sun.

Stellar transit near the cometary nucleus provides us a clue for reliable estimation of number density of dust particles in cometary coma. Direct measurement of the extinction of stellar flux makes it possible to infer opacity of coma in the region crossed by star. Then, based on realistic assumptions on cross section for extinction in dust particles, number density of dust particles can be retrieved from the optical depth of coma. However, the number density is a key characteristic in retrieval of reflectance (albedo) of dust particles from surface photometry of comets (Combes et al., 1983; Larson and A'Hearn, 1984). Furthermore, the number density is an important parameter for estimation of dust production by a comet. It is important that the distant comets with perihelion distance over 4 AU often reveal no significant gaseous emission (Korsun et al., 2010; Rousselot et al., 2014), allowing us to study dust in the distant comets with broadband filters at minimum risk of contamination of the signal with gaseous emission.

In this paper, we present and analyze the results of photometric observations of the distant comet C/2011 J2 (LINEAR) (hereafter C/2011 J2). Comet C/2011 J2 was discovered within the Lincoln Near-Earth Asteroid Research project as a 20-th magnitude small diffuse object on May 4.19, 2011 at heliocentric distance $r = 8.63$ AU and geocentric distance $\Delta$=8.02 AU. The comet has a perihelion distance of 3.44 AU, it passed through perihelion on December 25, 2013; an orbital inclination $i = 122.8°$ and eccentricity $e = 1.0005132$ (Williams, 2011). The comet got fragmented after the perihelion passage and two additional components (B and C) were detected. The breakup was first observed by Manzini et al. (Manzini, 2014) on the 27th of August 2014 and then confirmed by other observers.

## 2. Observations and reduction

Comet C/2011 J2 has been monitored with three different telescopes. The observations were performed with the 70-cm AZT-8 (observation station Lisnyky of the Astronomical Observatory of Taras Shevchenko National University, Ukraine), the 61-cm telescope at the Skalnate Pleso (AI SAS, Slovakia) and the 1-m Zeiss-1000 (SAO RAN, Russia). Main characteristics of telescopes and CCD detectors used are given in Table 1.

Tab.1. Equipment of the observations.

| Telescope | Diameter, [m] | CCD | Pixel size, [μm×μm] | Scale[b], [″/pix] | Field of view, [′×′] | Filter |
|---|---|---|---|---|---|---|
| AZT-8 | 0.70 | PL47-10 FLI | 13×13 | 0.96 | 16×16 | R |
| TSP[a] | 0.61 | SBIG-ST-10XME | 9×9 | 1.6 | 19×13 | R |
| Zeiss-1000 | 1.00 | EEV 42-40 | 13.5×13.5 | 0.48 | 8×8 | R |

[a] – TSP- Telescope at the Skalnate Pleso
[b] – value take into account the applied binning

We conducted photometric observations of the comet C/2011 J2 in 2013 and 2014 with the AZT-8 telescope at observation station Lisnyky of the Astronomical Observatory of Taras Shevchenko National University, Ukraine. These observations span the epochs before and after the perihelion passage. From May to November 2014, we also carried out observations of the comet with the 61-cm Telescope at the Skalnate Pleso (AI SAS, Slovakia). These observations were done after the perihelion passage. The latest photometric observations were made on 25 November 2014 with the Zeiss-1000 telescope (SAO RAS, Russia). The observations obtained in 2013 crowded field with seeing being stable around 1.9 arcsec and for 2014 near 1.8 arcsec.

All observations of the comet were performed with the R broadband filter of the Johnson-Cousin system. We applied 2×2 pixels (for images from AZT-8 and Zeiss-1000) and 3×3 pixels (TSP) binning to the photometric frames. The telescope was guided on non-sidereal rate to compensate the comet motion during the exposures. All the images were corrected for bias, and flat-fielding in the standard manner, using the IDL[1] routines. We obtained a set of exposures from the morning twilight sky through the filter to create a flat field, obtained dark and bias images. In order to increase the signal-to-noise ratio (S/N) and remove residual background star contribution, images of the comet obtained on each epoch were first aligned for the comet photometric center, the optocenter, and, then, they were treated to combine a median image. We attribute the comet optocenter to the central isophote that encloses the maximum of brightness of the comet. In order to calculate the sky background count, we adapt the routines developed by Landsman (1993). We also observed a field with standard stars (SA 104-330, SA 92-259) at different air masses to provide the information needed for the photometric reduction (Landolt, 1992). In this article we include only the observations obtained on photometric nights.

## 3. Results of observations

### 3.1 Morphology of comets

Fragmentation of a comet is plausibly accompanied with ejection of pristine materials well preserved inside the cometary nucleus. Therefore, in the case of C/2011 J2, we may have an opportunity to observe internal composition of its nucleus that did not undergo significant heating. Comet C/2011 J2 got fragmented after the perihelion passage in December 2013, the exact fragmentation date remains unknown. Two fragments (B, C) in vicinity of the C/2011 J2 nucleus were reported in the literature (Williams, 2014a; Williams, 2014b). However, we detected only the brighter component B, which was more active between September 19 and November 5, 2014.

In order to enhance the B component in the images of the dusty coma, we used the special software Cometary Coma Image Enhancement Facility[2] (Samarasinha, N. H., Martin, M. P., Larson, S. M, 2013) and Astroart[3], which are provided with a number of digital filters. In Fig. 1a, it is shown the initial image of the comet obtained on September 18, 2014, by co-adding basic frames. Division of the image in Fig. 1a by cometocentric distance ρ enhances contribution of outer coma. This procedure is aimed to compensate dimming of outer coma caused by decreasing volume concentration of dust particles. This is accomplished with the software Cometary Coma Image Enhancement Facility (Samarasinha, N. H., Martin, M. P., Larson, S. M, 2013). In addition, we subtract background using the IDL program. The image emerging from these operations is shown in Fig. 1b. In what follows, we convolve the image in Fig. 1b with the point spread function (PSF) and, then, apply the unsharp masking and the Gaussian blur procedures. It is done using the corresponding digital filters included in software Astroart. Note, similar processing has been applied to the images of Comet C/2011 J2 in Scarmato (2014 a,b). Also, the same technique was successfully applied in Kharchuk et al. (2015) for enhancing the structure of coma in Comet C/2012 S1 (ISON). The resulting image is demonstrated in Fig. 1c. The final image unambiguously reveals a companion (fragment B) in vicinity of the C/2011 J2 nucleus.

[1] http://www.exelisvis.com/ProductsServices/IDL.aspx

[2] http://www.psi.edu/research/cometimen

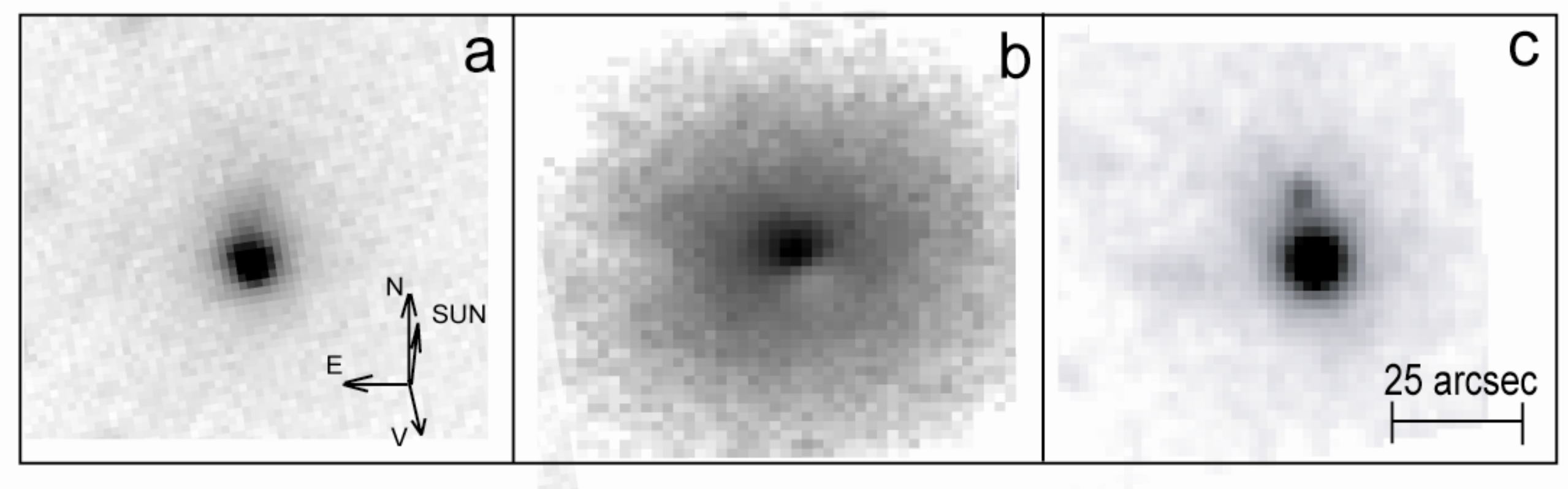

Fig.1. Morphological analysis of cropped R-filter images of comets C/2011 J2 (LINEAR) obtained on September 18, 2014. (a) – original images of the comet; (b) – the 1/ρ model using for cropped mages. (C) – final image after all using procedures that shows fragment B near cometary nucleus.

Figure 2 presents the enhanced images (for two epochs of observation), where showing the position of the main nucleus and the fragment B and their brightness profiles.

---

[3] http://www.msb-astroart.com/

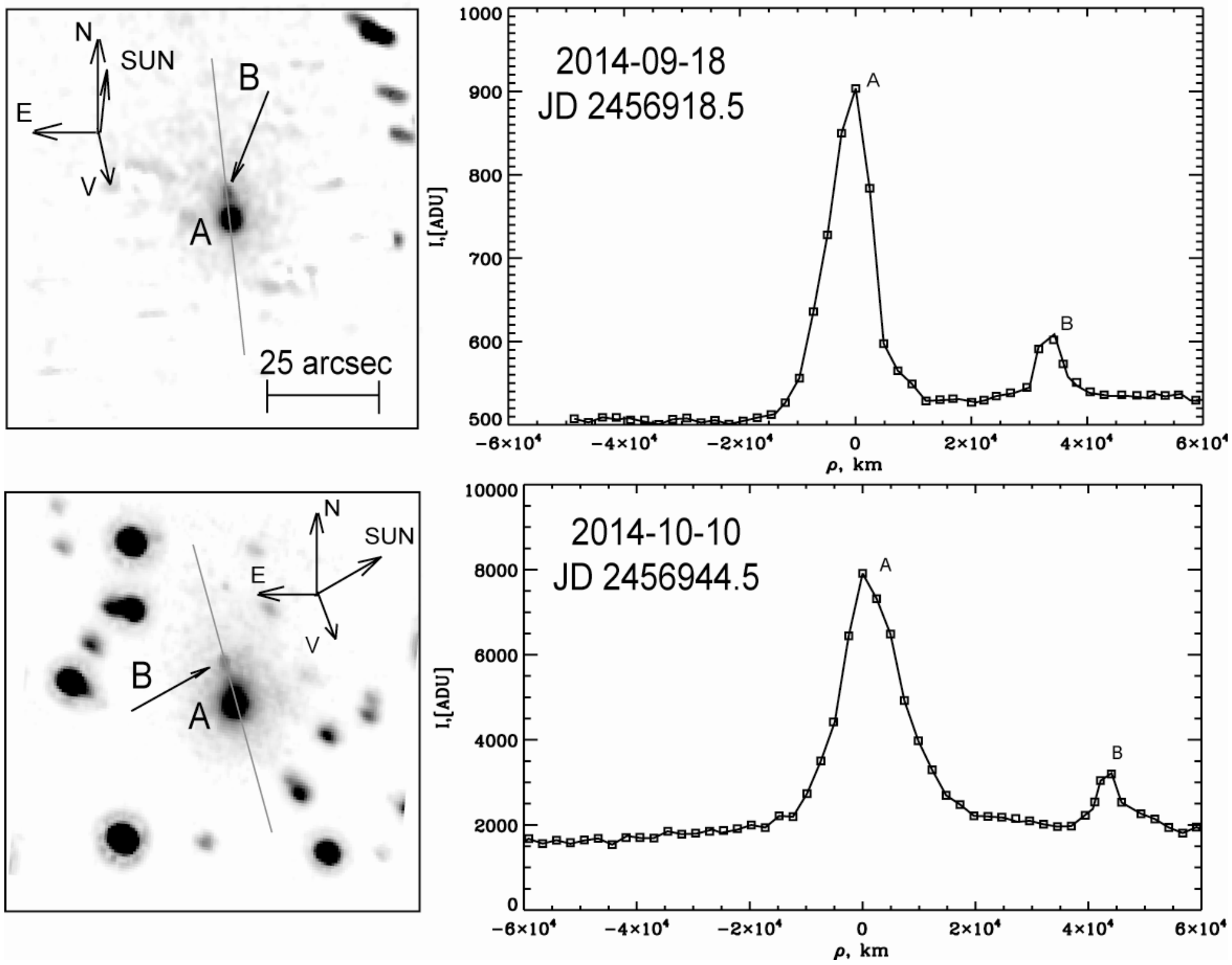

Fig.2. Images of Comet C/2011 J2 (LINEAR) (A) and component B on two epochs (left) and the corresponding brightness profiles of the comet with component B (right). The directions to the North (N), to the East (E), to the Sun and moving direction (V) of the comet are indicated.

We estimate the persistence E of fragment B for time period of observation from September 18 through October 5, 2014, using the equation adapted from Sekanina (1977; 1982):

$$E = 1.015 \cdot A_{sf} / \left[a \cdot (1+e) \cdot (1-e)\right]^{1/2} \quad (1)$$

where $A_{sf}$ is the length of the heliocentric arc of the orbit (in degrees) that the fragment passed through between the splitting event $T_s$ and final observation $T_f$, a=6454.884 AU is the semi major axis, and e=1.0005334is the eccentricity of the cometary orbit (JPL Small-body database[4]). Eq. (1) yields E=29.1 days.

We compute the deceleration parameter Γ=60±61 (10-5 solar units) based on the fragment endurance using the relationship (Sekanina, 1982; Boehnhardt, 2004)

[4] http://ssd.jpl.nasa.gov/

$$E(days) = 690(\pm 180) \cdot \Gamma^{-0.77\,(\pm 0.07)}, \qquad (2)$$

Thus, our estimation with eq. (2) suggests the component B is a short-lived one. We also estimate the relative speed of fragment B from our observations between 18 September and 10 October, 2014. We found the relative projected velocity of the fragment B is ~ 0.17″/day, or ~4.9 m/s. It is worth to note that, in other comets, the relative speed of the fragments shortly after their fragmentation spans the range from 0.1 m/s through 15 m/s; whereas, in vast majority of cases, the range appears to be squeezed to 0.3 – 4 m/s (Boehnhardt, 2004). Therefore, in sense of the split velocity, the C/2011 J2 breakup is consistent with what was found in other comets. However, this also could suggest common mechanism and/or circumstances of the breakup.

## 3.2 Photometry of comet

Using the images of Comet C/2011 J2 (LINEAR) obtained with the broadband R filter, we estimate the magnitude, the upper limit of effective radius of the nucleus, and the dust production rate. Taking into account the seeing of the stars at the night of observation, we perform photometry of three areas, all centered on the optocenter. The apertures were selected using the method actively exploited in a cometary photometry (in O'Ceallaigh et al., 1995; Licandro et al., 2000; Lowry and Fitzsimmons, 2005; Mazzotta Epifani et al., 2008; 2014). Fifty profiles have been centered on the cometary optocenter. They have been extracted and averaged to obtain a mean target profile that is shown on the right in Fig. 3. Normalized profile has been compared versus PSF stellar profile, which is obtained for the standard field stars; FWHM of the stars inferred from our observations spanned the range from 1.8 to 1.9 arcsec.

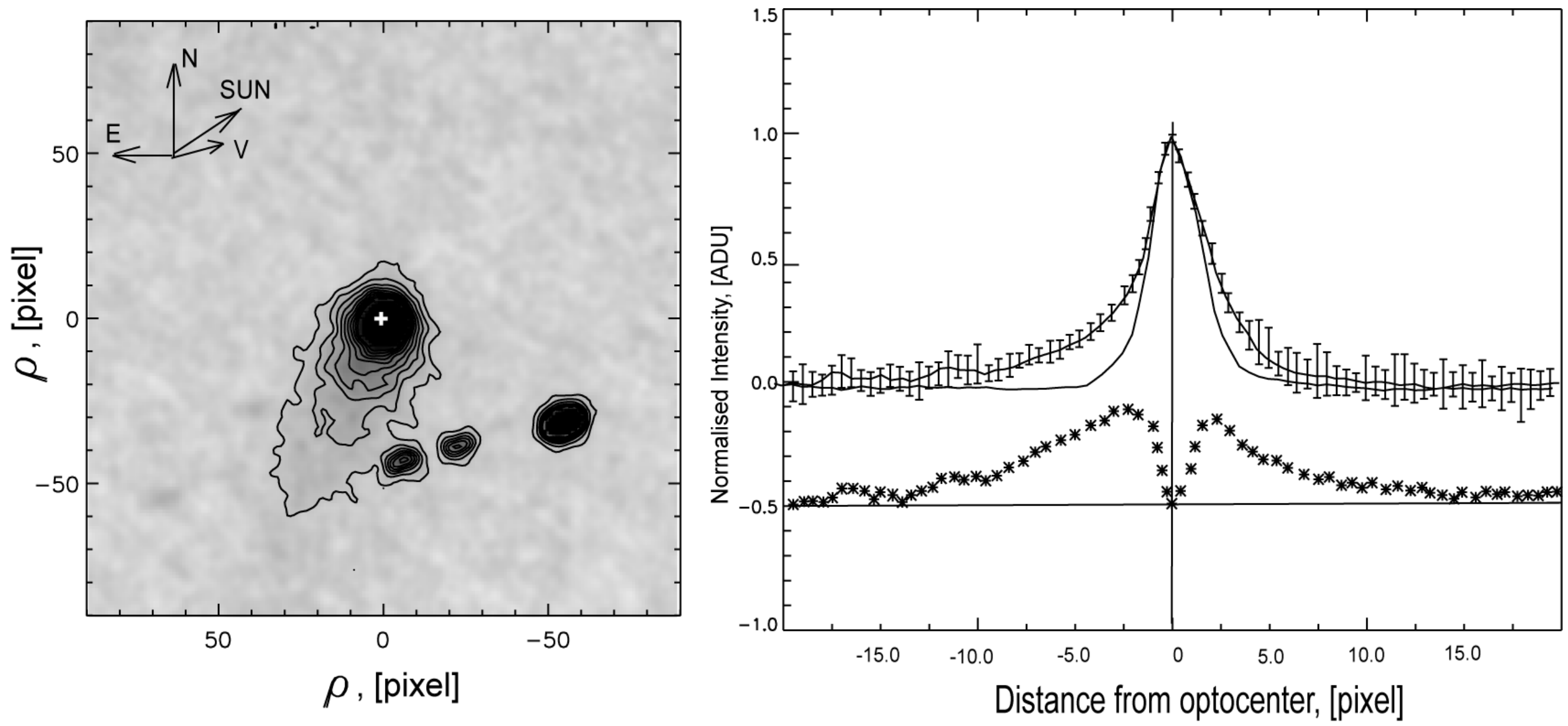

Fig.3. The image of the comet C/2011 J2 (LINEAR) together with contour plots, obtained on April 25, 2013 (left). The averaged profile (continuous line with error bars), average stellar PSF profile (continuous line) and subtraction of the stellar profile to the target profile (black asterisk).

Therefore, we select a circular aperture with diameter of ~ 2 arcsec that is approximately equal to the angular size of a star. In what follows, it is assumed that the flux measured with this aperture predominantly is originated from the nucleus; whereas, contribution from coma is weak. Evidently, this is a rough approximation for active comets, such as C/2011 J2. This implies that our estimation of the C/2011 J2 nucleus size corresponds to its upper limit. For calculation of the Afρ parameter and dust production rate, we used the circular apertures with radius from 4 arcsec to 14 arcsec.

We calculate magnitudes of the comet adjusted for r = Δ = 1 AU and phase angle α = 0º using the relationship as follows (Meech et al., 2009):

$$m_R(1,1,0) = m_R - 5 \cdot \log(r) - 2.5 \cdot \log(\Delta) - (\beta \cdot \alpha), \qquad (3)$$

where r and Δ are heliocentric and geocentric distances of the comet (in AU); $m_R$ is the R magnitude of the comet; α is the phase angle of the comet (in degree); β = 0.04 is the linear phase coefficient in magnitudes per degree (Snodgrass et al., 2008). The results are summarized in Table 2 and Fig.4.

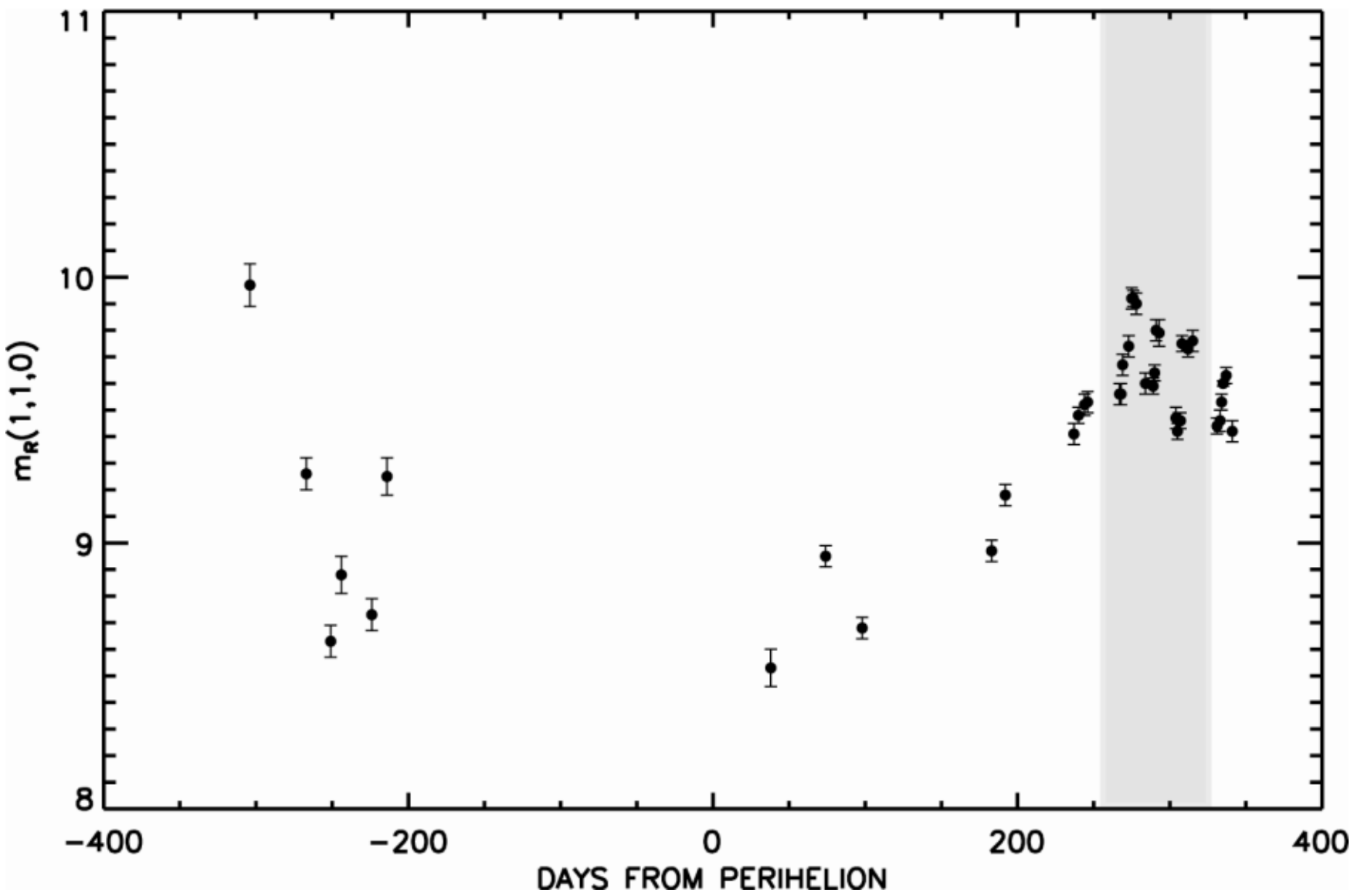

Fig.4. Time evolution of $m_R$ (1,1,0) of the Comet C/2011 J2 (LINEAR).

Tab.2 Log of the observations of the comet C/2011 J2 (LINEAR) and obtained in R filter.

| Date of obs. | Orb.[a] | r[b], (AU) | Δ[c], (AU) | Phase angle, (°) | $T_{exp}$, s × N | $m_R$ | aperture radius arcses | aperture radius km | $m_R$(1,1,0) | Afro, cm | Telescope |
|---|---|---|---|---|---|---|---|---|---|---|---|
| 24/02/2013<br>2456347.5 | I | 4.41 | 3.54 | 7.04 | 60×20 | 15.28±0.07<br>15.41±0.08<br>15.58±0.08<br>15.82±0.07<br>16.22±0.08<br>16.75±0.08 | 14<br>12<br>10<br>8<br>6<br>4 | 34507.1<br>29577.5<br>24648.0<br>19718.4<br>14788.8<br>9859.18 | 9.03±0.07<br>9.16±0.08<br>9.33±0.08<br>9.57±0.07<br>9.97±0.08<br>10.50±0.08 | 603±27<br>654±29<br>710±31<br>708±31<br>690±31<br>667±30 | AZT-8 |
| 02/04/2013<br>2456384.5 | I | 4.21 | 3.75 | 12.9 | 60×15 | 14.86±0.05<br>14.97±0.05<br>15.15±0.06<br>15.38±0.06<br>15.77±0.06<br>16.28±0.06 | 14<br>12<br>10<br>8<br>6<br>4 | 36554.2<br>31332.2<br>26110.1<br>20888.1<br>15666.1<br>10444.0 | 8.35±0.05<br>8.46±0.05<br>8.64±0.06<br>8.87±0.06<br>9.26±0.06<br>9.77±0.06 | 899 ±40<br>959±43<br>1030±46<br>1019±45<br>1002±45<br>950±42 | AZT-8 |
| 18/04/2013<br>2456400.5 | I | 4.13 | 3.91 | 13.9 | 120×10 | 14.40±0.07<br>14.47±0.07<br>14.62±0.07<br>14.85±0.06<br>15.23±0.06<br>15.82±0.07 | 14<br>12<br>10<br>8<br>6<br>4 | 38113.8<br>32669.0<br>27224.2<br>21779.3<br>16334.5<br>10889.7 | 7.80±0.07<br>7.87±0.07<br>8.02±0.07<br>8.25±0.06<br>8.63±0.06<br>9.22±0.07 | 1399±62<br>1600±72<br>1703±76<br>1684±75<br>1611±72<br>1473±66 | AZT-8 |
| 25/04/2013<br>2456407.5 | I | 4.09 | 3.99 | 14.2 | 60×15 | 14.70±0.07<br>14.76±0.07<br>14.93±0.07<br>15.14±0.07<br>15.52±0.07<br>16.13±0.07 | 14<br>12<br>10<br>8<br>6<br>4 | 38893.6<br>33337.4<br>27781.2<br>22224.9<br>16668.7<br>11112.5 | 8.06±0.07<br>8.12±0.07<br>8.29±0.07<br>8.50±0.07<br>8.88±0.07<br>9.49±0.07 | 1051±47<br>1229±55<br>1308±58<br>1270±57<br>1237±55<br>1121±50 | AZT-8 |
| 15/05/20132<br>456427.5 | I | 4.00 | 4.21 | 13.8 | 120×10 | 14.55±0.06<br>14.66±0.06<br>14.81±0.05<br>15.05±0.05<br>15.42±0.06<br>15.96±0.06 | 14<br>12<br>10<br>8<br>6<br>4 | 41038.2<br>35175.6<br>29313.0<br>23450.4<br>17587.8<br>11725.2 | 7.86±0.06<br>7.97±0.06<br>8.12±0.05<br>8.36±0.05<br>8.73±0.06<br>9.27±0.06 | 1243±55<br>1363±61<br>1437±64<br>1434±64<br>1372±61<br>1302±58 | AZT-8 |
| 25/05/2013 | I | 3.96 | 4.32 | 13.2 | 60×25 | 15.05±0.07 | 14 | 42110.4 | 8.35±0.07 | 785±35 | AZT-8 |
| 2456437.5 | | | | | | 15.18±0.08<br>15.37±0.07<br>15.63±0.07<br>15.95±0.07<br>16.43±0.08 | 12<br>10<br>8<br>6<br>4 | 36094.6<br>30078.9<br>24063.1<br>18047.3<br>12031.5 | 8.48±0.08<br>8.67±0.07<br>8.93±0.07<br>9.25±0.07<br>9.73±0.08 | 814±36<br>820±36<br>833±37<br>827±37<br>799±35 | |
| 01/02/2014<br>2456689.5 | O | 3.46 | 3.37 | 16.5 | 120×17 | 13.37±0.06<br>13.47±0.06<br>13.80±0.06<br>14.06±0.06<br>14.53±0.07<br>15.30±0.09 | 14<br>12<br>10<br>8<br>6<br>4 | 32850.0<br>28157.2<br>23464.3<br>18771.4<br>14078.6<br>9385.72 | 7.37±0.06<br>7.47±0.06<br>7.80±0.06<br>8.06±0.06<br>8.53±0.07<br>9.30±0.09 | 843±37<br>1132±50<br>1267±57<br>1351±60<br>1300±58<br>1275±57 | AZT-8 |
| 09/03/2014<br>2456725.5 | O | 3.51 | 3.80 | 14.9 | 60×20 | 14.25±0.03<br>14.38±0.03<br>14.58±0.03<br>14.82±0.04<br>15.18±0.04<br>15.75±0.07 | 14<br>12<br>10<br>8<br>6<br>4 | 37041.6<br>31749.9<br>26458.3<br>21166.6<br>15875.0<br>10583.3 | 8.02±0.03<br>8.15±0.03<br>8.35±0.03<br>8.59±0.04<br>8.95±0.04<br>9.52±0.03 | 1135±51<br>1201±54<br>1255±56<br>1272±57<br>1254±56<br>1160±52 | AZT-8 |
| 02/04/2014<br>2456749.5 | O | 3.56 | 4.15 | 12.1 | 60×15 | 14.16±0.03<br>14.29±0.03<br>14.47±0.03<br>14.69±0.03<br>15.02±0.04<br>15.51±0.04 | 14<br>12<br>10<br>8<br>6<br>4 | 40453.3<br>34674.2<br>28895.2<br>23116.2<br>17337.1<br>11558.1 | 7.82±0.03<br>7.95±0.03<br>8.13±0.03<br>8.35±0.03<br>8.68±0.04<br>9.17±0.04 | 1466±65<br>1535±69<br>1560±70<br>1528±68<br>1503±67<br>1365±61 | AZT-8 |
| 21/06/2014<br>2456829.5 | O | 3.81 | 4.22 | 13.2 | 180×11 | 14.18±0.03<br>14.23±0.03<br>14.39±0.03<br>14.44±0.04<br>15.53±0.04<br>15.71±0.05 | 14<br>12<br>10<br>8<br>6<br>4 | 68559.4<br>58765.2<br>48971.0<br>39176.8<br>29382.6<br>19588.4 | 7.62±0.03<br>7.67±0.03<br>7.83±0.03<br>7.88±0.04<br>8.97±0.04<br>9.15±0.05 | 1380±62<br>1417±63<br>1588±71<br>1616±72<br>1550±69<br>1461±65 | TSP |
| 05/07/2014<br>2456843.5 | O | 3.86 | 4.11 | 14.2 | 180×9 | 14.21±0.03<br>14.38±0.03<br>14.47±0.03<br>14.66±0.03<br>15.76±0.04<br>15.81±0.04 | 14<br>12<br>10<br>8<br>6<br>4 | 66772.3<br>57233.4<br>47694.5<br>38155.6<br>28616.7<br>19077.8 | 7.63±0.03<br>7.80±0.03<br>7.89±0.03<br>8.08±0.03<br>9.18±0.04<br>9.23±0.04 | 1295±58<br>1404±63<br>1568±70<br>1580±71<br>1412±63<br>1366±61 | TSP |
| 19/08/2014<br>2456888.5 | O | 4.06 | 3.65 | 13.7 | 60×15 | 14.90±0.03<br>15.04±0.04<br>15.24±0.04<br>15.46±0.03<br>15.82±0.04<br>16.32±0.04 | 14<br>12<br>10<br>8<br>6<br>4 | 35579.4<br>30496.6<br>25413.9<br>20331.1<br>15248.3<br>10165.5 | 8.49±0.03<br>8.63±0.04<br>8.83±0.04<br>9.05±0.03<br>9.41±0.04<br>9.91±0.04 | 1018±45<br>1075±48<br>1123±50<br>1101±49<br>1103±49<br>1075±48 | AZT-8 |
| 22/08/2014<br>2456891.5 | O | 4.08 | 3.63 | 13.5 | 180×10 | 14.91±0.03<br>15.11±0.04<br>15.30±0.04<br>15.52±0.03<br>15.88±0.03<br>16.41±0.04 | 14<br>12<br>10<br>8<br>6<br>4 | 58974.1<br>50549.2<br>42124.3<br>33699.5<br>25274.6<br>16849.7 | 8.51±0.03<br>8.71±0.04<br>8.90±0.04<br>9.12±0.03<br>9.48±0.03<br>10.01±0.04 | 735±33<br>798±35<br>834±37<br>817±36<br>811±36<br>844±37 | TSP |
| 26/08/2014<br>2456895.5 | O | 4.10 | 3.60 | 13.1 | 60×15 | 14.92±0.04<br>15.07±0.04<br>15.24±0.04<br>15.52±0.04<br>15.89±0.04<br>16.46±0.06 | 14<br>12<br>10<br>8<br>6<br>4 | 35092.0<br>30078.9<br>25065.7<br>20052.6<br>15039.4<br>10026.3 | 8.55±0.04<br>8.70±0.04<br>8.87±0.04<br>9.15±0.04<br>9.52±0.04<br>10.09±0.06 | 899±40<br>1013±45<br>1046±47<br>1106±49<br>1078±48<br>1061±47 | AZT-8 |
| 28/08/2014<br>2456897.5 | O | 4.11 | 3.58 | 12.8 | 180×10 | 14.85±0.04<br>15.09±0.04<br>15.23±0.04<br>15.48±0.04<br>15.89±0.04<br>16.16±0.05 | 14<br>12<br>10<br>8<br>6<br>4 | 58161.8<br>49852.9<br>41544.1<br>33235.3<br>24926.5<br>16617.6 | 8.49±0.04<br>8.73±0.04<br>8.87±0.04<br>9.12±0.04<br>9.53±0.04<br>9.80±0.05 | 626±28<br>792±35<br>866±38<br>872±39<br>827±37<br>884±39 | TSP |
| 18/09/2014<br>2456918.5 | O | 4.21 | 3.49 | 10.3 | 60×20/<br>180×15 | 14.82±0.04<br>15.03±0.03<br>15.25±0.04<br>15.49±0.04<br>15.81±0.04<br>16.28±0.04 | 14<br>12<br>10<br>8<br>6<br>4 | 34019.8<br>29159.8<br>24299.8<br>19439.9<br>14579.9<br>9719.93 | 8.57±0.04<br>8.78±0.03<br>9.00±0.04<br>9.24±0.04<br>9.56±0.04<br>10.03±0.04 | 849±38<br>873±39<br>879±39<br>877±39<br>895±40<br>931±41 | AZT-8 |
| | | | | | | *19.49±0.03 | 2 | 4859.96 | *13.24±0.03 | 88±14 | |
| 19/09/2014<br>2456919.5 | O | 4.22 | 3.48 | 10.2 | 180×15 | 14.93±0.03<br>15.21±0.03<br>15.38±0.03<br>15.59±0.04<br>15.81±0.04<br>16.38±0.04 | 14<br>12<br>10<br>8<br>6<br>4 | 56537.1<br>48460.4<br>40383.7<br>32306.9<br>24230.2<br>16153.5 | 8.68±0.03<br>8.96±0.03<br>9.13±0.03<br>9.34±0.04<br>9.56±0.04<br>10.13±0.04 | 775±34<br>873±39<br>802±36<br>779±35<br>759±34<br>842±37 | TSP |
| 20/09/2014<br>2456920.5 | O | 4.22 | 3.48 | 10.1 | 180×15 | 15.03±0.03<br>15.26±0.03<br>15.41±0.03<br>15.64±0.04<br>15.91±0.04<br>16.41±0.05 | 14<br>12<br>10<br>8<br>6<br>4 | 56537.1<br>48460.4<br>40383.7<br>32306.9<br>24230.2<br>16153.5 | 8.79±0.03<br>9.02±0.03<br>9.17±0.03<br>9.40±0.04<br>9.67±0.04<br>10.17±0.05 | 754±33<br>797±35<br>766±34<br>757±34<br>725±32<br>768±34 | TSP |
| 24/09/2014<br>2456924.5 | O | 4.24 | 3.48 | 9.7 | 180×12 | 15.16±0.03<br>15.38±0.04<br>15.50±0.04<br>15.72±0.04<br>15.98±0.04<br>16.48±0.04 | 14<br>12<br>10<br>8<br>6<br>4 | 56537.1<br>48460.4<br>40383.7<br>32306.9<br>24230.2<br>16153.5 | 8.92±0.03<br>9.14±0.04<br>9.26±0.04<br>9.48±0.04<br>9.74±0.04<br>10.24±0.04 | 713±32<br>754±33<br>718±32<br>704±31<br>655±29<br>688±30 | TSP |
| 26/09/2014<br>2456926.5 | O | 4.25 | 3.48 | 9.5 | 30×26 | 15.25±0.04<br>15.40±0.05<br>15.59±0.04<br>15.83±0.04<br>16.15±0.04<br>16.62±0.05 | 14<br>12<br>10<br>8<br>6<br>4 | 33922.3<br>29076.2<br>24230.2<br>19384.2<br>14538.1<br>9692.08 | 9.02±0.04<br>9.17±0.05<br>9.36±0.04<br>9.60±0.04<br>9.92±0.04<br>10.39±0.05 | 633±28<br>650±29<br>655±29<br>654±29<br>649±29<br>639±28 | AZT-8 |
| | | | | | | *19.51±0.05 | 2 | 9692.08 | *13.28±0.05 | 88±14 | |
| 28/09/2014**<br>2456928.5 | O | 4.26 | 3.49 | 9.5 | 180×18 | 15.23±0.03<br>15.45±0.03<br>15.62±0.04<br>15.86±0.04<br>16.17±0.03<br>16.63±0.04 | 14<br>12<br>10<br>8<br>6<br>4 | 56699.6<br>48599.7<br>40499.7<br>32399.8<br>24299.8<br>16199.9 | 8.98±0.03<br>9.20±0.03<br>9.37±0.04<br>9.61±0.03<br>9.92±0.03<br>10.38±0.04 | 629±28<br>641±28<br>639±28<br>638±28<br>622±27<br>653±29 | TSP |
| 29/09/2014<br>2456929.5 | O | 4.26 | 3.59 | 9.4 | 180×15 | 15.24±0.03<br>15.43±0.03<br>15.60±0.03<br>15.89±0.03<br>16.20±0.04<br>16.60±0.04 | 14<br>12<br>10<br>8<br>6<br>4 | 58324.2<br>49992.2<br>41660.2<br>33328.1<br>24996.1<br>16664.1 | 8.94±0.03<br>9.13±0.03<br>9.30±0.03<br>9.59±0.03<br>9.90±0.04<br>10.30±0.04 | 665±29<br>641±28<br>640±28<br>667±30<br>648±29<br>663±29 | TSP |
| 05/10/2014<br>2456935.5 | O | 4.30 | 3.51 | 9.1 | 180×10 | 15.17±0.03<br>15.26±0.03<br>15.41±0.03<br>15.60±0.04<br>15.86±0.04<br>16.31±0.04 | 14<br>12<br>10<br>8<br>6<br>4 | 57024.5<br>48878.2<br>40731.8<br>32585.4<br>24439.1<br>16292.7 | 8.91±0.03<br>9.00±0.03<br>9.15±0.03<br>9.34±0.04<br>9.60±0.04<br>10.05±0.04 | 876±39<br>883±39<br>842±37<br>802±36<br>768±34<br>715±32 | TSP |
| | | | | | | *19.57±0.04 | 2 | 5091.4 | 13.2±0.04 | 86±14 | |
| 10/10/2014<br>2456940.5 | O | 4.32 | 3.54 | 9.1 | 60×20 | 15.11±0.04<br>15.21±0.04<br>15.37±0.03<br>15.57±0.03<br>15.88±0.04<br>16.36±0.04 | 14<br>12<br>10<br>8<br>6<br>4 | 34507.1<br>29577.5<br>24648.0<br>19718.4<br>14788.8<br>9859.18 | 8.82±0.04<br>8.92±0.04<br>9.08±0.03<br>9.28±0.03<br>9.59±0.04<br>10.07±0.04 | 848±38<br>880±39<br>878±39<br>845±38<br>816±36<br>767±34 | AZT-8 |
| | | | | | | *19.62±0.04 | 2 | 9859.18 | *13.33±0.04 | 82± 12 | |
| 11/10/2014 | O | 4.33 | 3.55 | 9.1 | 180×10 | 15.17±0.03 | 14 | 57674.4 | 8.87±0.03 | 692 31 | TSP |
| 2456941.5 | | | | | | 15.31±0.03<br>15.57±0.03<br>15.75±0.03<br>15.94±0.03<br>16.58±0.04 | 12<br>10<br>8<br>6<br>4 | 49435.2<br>41196.0<br>32956.8<br>24717.6<br>16478.4 | 9.01±0.03<br>9.27±0.03<br>9.45±0.03<br>9.64±0.03<br>10.28±0.04 | 832 37<br>743 33<br>702 31<br>743 33<br>725 32 | |
| 12/10/2014<br>2456942.5 | O | 4.34 | 3.56 | 9.1 | 60×18 | 15.24±0.04<br>15.39±0.04<br>15.57±0.04<br>15.79±0.04<br>16.11±0.04<br>16.59±0.05 | 14<br>12<br>10<br>8<br>6<br>4 | 34702.1<br>29744.7<br>24787.2<br>19829.8<br>14872.3<br>9914.88 | 8.93±0.04<br>9.08±0.04<br>9.26±0.04<br>9.48±0.04<br>9.80±0.04<br>10.28±0.05 | 693±31<br>718±32<br>724±32<br>709±31<br>697±31<br>686±30 | AZT-8 |
| | | | | | | *19.64±0.05 | 2 | 9914.88 | *13.33±0.05 | 83± 12 | |
| 14/10/2014<br>2456944.5 | O | 4.34 | 3.56 | 9.1 | 60×20 | 15.16±0.05<br>15.31±0.05<br>15.49±0.05<br>15.72±0.05<br>16.10±0.05<br>16.68±0.07 | 14<br>12<br>10<br>8<br>6<br>4 | 34702.1<br>29744.7<br>24787.2<br>19829.8<br>14872.3<br>9914.88 | 8.85±0.05<br>9.00±0.05<br>9.18±0.05<br>9.41±0.05<br>9.79±0.05<br>10.37±0.07 | 644±28<br>732±32<br>779±35<br>771±34<br>758±34<br>746±33 | AZT-8 |
| | | | | | | *19.68±0.07 | 2 | 9914.88 | *13.37±0.07 | 80± 12 | |
| 25/10/2014<br>2456955.5 | O | 4.35 | 3.57 | 9.2 | 60×25 | 14.88±0.04<br>15.03±0.04<br>15.22±0.04<br>15.44±0.04<br>15.80±0.05<br>16.38±0.06 | 14<br>12<br>10<br>8<br>6<br>4 | 34799.6<br>29828.2<br>24856.8<br>19885.5<br>14914.1<br>9942.74 | 8.55±0.04<br>8.70±0.04<br>8.89±0.04<br>9.11±0.04<br>9.47±0.05<br>10.05±0.06 | 896±40<br>1019±45<br>1064±47<br>1043±46<br>1035±46<br>1019±45 | AZT-8 |
| | | | | | | *19.76±0.06 | 2 | 9942.74 | *13.44±0.06 | 75±12 | |
| 26/10/2014<br>2456956.5 | O | 4.35 | 3.58 | 9.2 | 180×10 | 14.86±0.03<br>15.10±0.03<br>15.24±0.03<br>15.48±0.03<br>15.75±0.04<br>16.31±0.05 | 14<br>12<br>10<br>8<br>6<br>4 | 58161.8<br>49852.9<br>41544.1<br>33235.3<br>24926.5<br>16617.6 | 8.53±0.03<br>8.77±0.03<br>8.91±0.03<br>9.15±0.03<br>9.42±0.04<br>9.98±0.05 | 904±40<br>1009±45<br>970±43<br>968±43<br>918±41<br>982±44 | TSP |
| 27/10/2014<br>2456957.5 | O | 4.36 | 3.59 | 9.2 | 180×10 | 14.87±0.03<br>15.09±0.03<br>15.27±0.03<br>15.49±0.03<br>15.81±0.04<br>16.37±0.04 | 14<br>12<br>10<br>8<br>6<br>4 | 58324.2<br>49992.2<br>41660.2<br>33328.1<br>24996.1<br>16664.1 | 8.52±0.03<br>8.74±0.03<br>8.92±0.03<br>9.14±0.03<br>9.46±0.04<br>10.02±0.04 | 864v38<br>965±43<br>971±43<br>952±42<br>936±42<br>983±44 | TSP |
| 28/10/2014<br>2456958.5 | O | 4.36 | 3.60 | 9.3 | 180×10 | 14.95±0.03<br>15.16±0.03<br>15.28±0.03<br>15.49±0.03<br>15.82±0.03<br>16.39±0.04 | 14<br>12<br>10<br>8<br>6<br>4 | 58486.7<br>50131.4<br>41776.2<br>33421.0<br>25065.7<br>16710.5 | 8.59±0.03<br>8.80±0.03<br>8.92±0.03<br>9.13±0.03<br>9.46±0.03<br>10.03±0.04 | 848±38<br>956±43<br>971±43<br>943±42<br>878±39<br>913±41 | TSP |
| 29/10/2014<br>2456959.5 | O | 4.37 | 3.60 | 9.3 | 180×10 | 15.12±0.03<br>15.29±0.03<br>15.44±0.03<br>15.69±0.03<br>16.11±0.04<br>16.67±0.04 | 14<br>12<br>10<br>8<br>6<br>4 | 58486.7<br>50131.4<br>41776.2<br>33421.0<br>25065.7<br>16710.5 | 8.76±0.03<br>8.93±0.03<br>9.08±0.03<br>9.33±0.03<br>9.75±0.04<br>10.31±0.04 | 695±31<br>776±34<br>857±38<br>863±38<br>825±37<br>827±37 | TSP |
| 02/11/2014<br>2456963.5 | O | 4.45 | 3.80 | 10.4 | 180×15 | 15.52±0.03<br>15.67±0.03<br>15.84±0.03<br>16.00±0.03<br>16.29±0.04<br>16.80±0.04 | 14<br>12<br>10<br>8<br>6<br>4 | 61735.9<br>52916.5<br>44097.1<br>35277.7<br>26458.3<br>17638.8 | 8.96±0.03<br>9.11±0.03<br>9.28±0.03<br>9.44±0.03<br>9.73±0.04<br>10.24±0.04 | 639±28<br>681±30<br>668± 30<br>619±27<br>603±27<br>593±26 | TSP |
| 05/11/2014<br>2456966.5 | O | 4.47 | 3.85 | 10.7 | 180×10 | 15.58±0.05<br>15.72±0.05<br>15.90±0.05<br>16.06±0.04<br>16.37±0.05<br>16.94±0.06 | 14<br>12<br>10<br>8<br>6<br>4 | 37529.0<br>32167.7<br>26806.4<br>21445.1<br>16083.8<br>10722.6 | 8.97±0.05<br>9.11±0.05<br>9.29±0.05<br>9.45±0.04<br>9.76±0.05<br>10.33±0.06 | 576±25<br>649±29<br>647±29<br>600±27<br>590±26<br>576±25 | AZT-8 |
| | | | | | | *19.87±0.06 | 2 | 10722.6 | *13.26±0.06 | 77 3 | |
| 21/11/2014<br>2456982.5 | O | 4.55 | 4.13 | 11.8 | 60×17/<br>180×10 | 15.31±0.03<br>15.44±0.03<br>15.62±0.03<br>15.91±0.04<br>16.27±0.04<br>16.83±0.04 | 14<br>12<br>10<br>8<br>6<br>4 | 40258.3<br>34507.1<br>28755.9<br>23004.8<br>17253.6<br>11502.4 | 8.43±0.03<br>8.60±0.03<br>8.78±0.03<br>9.06±0.04<br>9.42±0.04<br>10.01±0.04 | 700±31<br>781±35<br>839±37<br>853±38<br>839±37<br>833±37 | AZT-8 |
| 23/11/2014<br>2456984.5 | O | 4.57 | 4.17 | 11.9 | 180×10 | 15.28±0.03<br>15.51±0.03<br>15.72±0.04<br>16.01±0.04<br>16.34±0.04<br>16.87±0.04 | 14<br>12<br>10<br>8<br>6<br>4 | 67747.1<br>58068.9<br>48390.8<br>38712.6<br>29034.5<br>19356.3 | 8.40±0.03<br>8.63±0.03<br>8.84±0.04<br>9.13±0.04<br>9.46±0.04<br>9.99±0.04 | 693±31<br>753±33<br>766±34<br>800±36<br>809±36<br>857±38 | TSP |
| 24/11/2014<br>2456985.5 | O | 4.57 | 4.19 | 11.9 | 180×10 | 15.32±0.03<br>15.57±0.03<br>15.78±0.03<br>16.13±0.03<br>16.42±0.04<br>16.90±0.04 | 14<br>12<br>10<br>8<br>6<br>4 | 68072.0<br>58347.4<br>48622.9<br>38898.3<br>29173.7<br>19449.1 | 8.43±0.03<br>8.68±0.03<br>8.89±0.03<br>9.24±0.03<br>9.53±0.04<br>10.01±0.04 | 678±30<br>703±31<br>689±31<br>761±34<br>769±34<br>830± 37 | TSP |
| 25/11/2014<br>2456986.5 | O | 4.58 | 4.21 | 11.9 | 180×10<br>/<br>300×15 | 15.36±0.01<br>15.64±0.01<br>15.80±0.01<br>16.22±0.01<br>16.51±0.01<br>16.94±0.02 | 14<br>12<br>10<br>8<br>6<br>4 | 20519.1<br>17587.8<br>14656.5<br>11725.2<br>8793.89<br>5862.59 | 8.45±0.01<br>8.73±0.01<br>8.89±0.01<br>9.31±0.01<br>9.60±0.01<br>10.03±0.02 | 670±30<br>664±29<br>650±29<br>766±34<br>740±33<br>820±36 | Zeiss-1000 |
| 27/11/2014<br>2456988.5 | O | 4.59 | 4.26 | 12.0 | 180×10 | 15.50±0.03<br>15.69±0.03<br>15.91±0.03<br>16.25±0.03<br>16.57±0.03<br>17.00±0.03 | 14<br>12<br>10<br>8<br>6<br>4 | 69209.2<br>59322.2<br>49435.2<br>39548.1<br>29661.1<br>19774.1 | 8.56±0.03<br>8.75±0.03<br>8.97±0.03<br>9.313±0.03<br>9.63 ±0.03<br>10.06±0.03 | 634±28<br>628±28<br>632±28<br>692±31<br>706±31<br>721±32 | TSP |
| 01/12/2014<br>2456992.5 | O | 4.61 | 4.35 | 12.1 | 60×25 | 15.55±0.03<br>15.70±0.03<br>15.92±0.04<br>16.13±0.04<br>16.40±0.04<br>17.04±0.04 | 14<br>12<br>10<br>8<br>6<br>4 | 70671.4<br>60575.5<br>50479.6<br>40383.7<br>30287.7<br>20191.8 | 8.55±0.03<br>8.71±0.03<br>8.93±0.04<br>9.13±0.04<br>9.43±0.04<br>10.06±0.04 | 609±27<br>745±33<br>744±33<br>696±31<br>703±31<br>699±31 | TSP |

[a]Orbital arc: I - pre-perihelion; O - post-perihelion
[b] r is heliocentric distance
[c] Δ is geocentric distance.
[*] estimated for fragment B
[**] day of stellar transit near cometary nucleus observation

### 3.2.1 Radius of the nucleus

We estimate the upper limit of the C/2011 J2 radius $R_N$ using the expression from Russell (1916), and adopted variant presented in Jewitt (1991) given by

$$p \cdot C_d \leq 2.25 \cdot 10^{22} \cdot \pi \cdot r^2 \cdot \Delta^2 \cdot 10^{-0.4(m_{SUN}-m_R)} / \Phi(\alpha), \qquad (4)$$

where $p$ is the geometric albedo; $\Phi(\alpha) = 10^{-0.4 \cdot \alpha \cdot \beta}$ is the phase function approximated at small phase angles; $\alpha$ is phase angle in degrees; . is a constant in$\psi$mag$\cdot$deg$^{-1}$; $C_d$ is the geometrical cross-section of the nucleus measured in m$^2$; $m_{SUN}$ = -27.094 is the magnitude of the Sun in R band (Holmberg et al., 2006). For computation we use the co-added image of the comet obtained on February 24, 2013, when the comet was not very active. With in the circular aperture ~ 2 arcsec, we infer the apparent magnitude of the comet $m_R$=18.75±0.04. We consider a range of geometric albedo $p$, from classical value 0.04 adopted for Jupiter-family comets and distant comets (Jorda et al., 2000; Meech et al., 2009) upto p = 0.13, the value measured in active Centaurs (Cruikshank and Brown, 1983; Bus et al. 1989; Jewitt, 2009). We adapt the phase coefficient β from 0.035 mag/deg (Snodgrass et al. 2011) to 0.11 (Peixinho et al., 2004). Fig. 5 demonstrates results of estimation of the C/2011 J2 nucleus size.

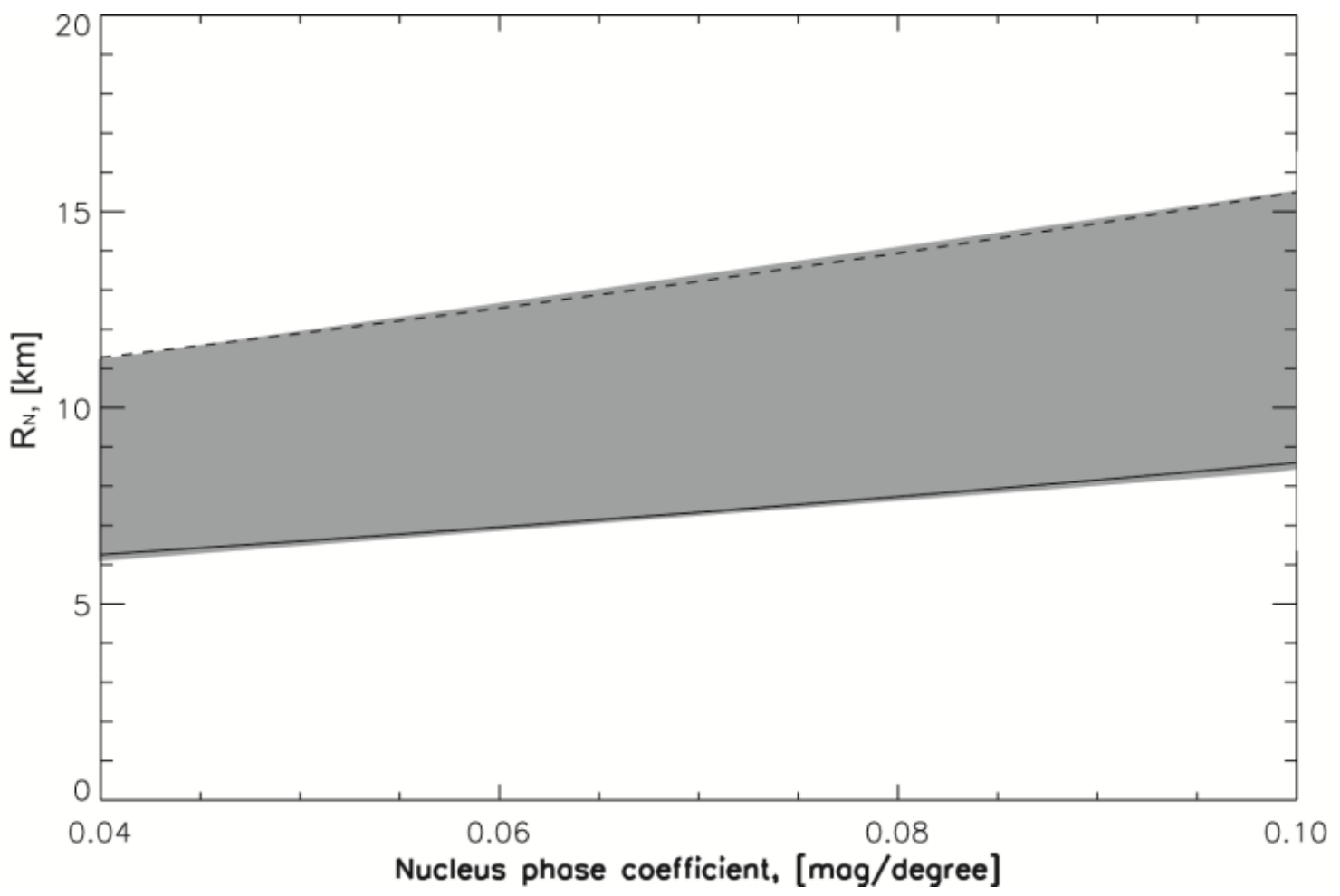

Fig. 5. The value of nucleus size of the comet C/2011 J2 (LINEAR). The grey area indicates the range of plausible nucleus radii.

To estimate the upper limits for the nucleus size in comet C/2011 J2, we consider all possible combinations of the input parameters in eq. (4), and constrain the C/2011 J2 radius to $R_N \leq 9$ km. We note the estimations of nuclei sizes in various short-period and long-period comets summarized in Meech et al. (2009). According to this survey, radius of nucleus in short-period comets spans the range from 0.6 km to 13.7 km; this estimation, however, is dependent on the method applied (PSF fit, Thermal, Distant). Radius of nucleus in long-period and dynamically new comets varies from 0.37 km to 56 km (Thermal, Distant methods). The estimations of nucleus radius in Centaurs reveal the range from 3.9 km to 51.4 km (Jewitt, 2009; Perna et al., 2010). It is clear that the size of Comet C/2011 J2 appears consistent with all these three groups of objects.

3.2.2 Afρ parameter

We estimate the Afρ parameter that is an aperture-independent characteristic in case a steady state of cometary coma (A'Hearn et al. 1984). The results [measured in cm] for a fixed aperture of ~25 000 km are presented in Fig. 6 as a function of the date of observation. In Fig. 6, the data obtained within the CARA Project[5] (private information from Bryssink E. and Milani G., 2015) are also plotted. As one can see here, in close epochs both results appear in good accordance to one another within the error.

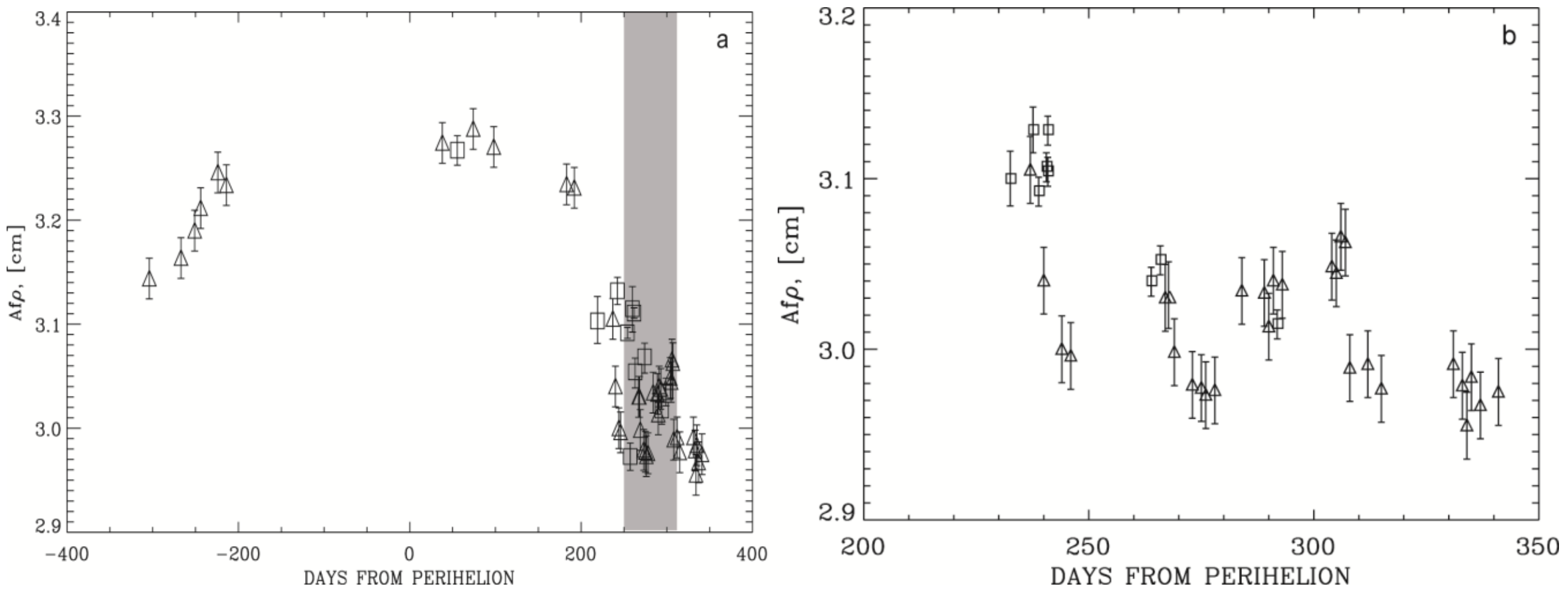

Fig.6 Time evolution of Afρ parameter in Comet C/2011 J2 (LINEAR). Triangle and square symbols present results obtained by us and the CARA team, respectively. The grey-shaded band in panel (a)

[5]cara.uai.it

shows the period when the fragmentation of the comet was apparent. Variation of the Afρ parameter over that time period can be seen in detail in panel (b).

The Afρ parameter was also estimated in vicinity of the fragment B. Using the aperture radius of 2 arcsec, it spans the range between 98 and 84 cm. When comparing the results for different active objects at the same heliocentric distance considered in Mazzotta Epifani et al. (2014) (shown in Fig. 4), Comet C/2011 J2 reveals similarity with the long-period comets and Centaurs.

In Fig. 7 we present the Afρ parameter for Comet C/2011 J2 as function of the aperture radius on several epochs of the observations; they correspond to the beginning of observation run, period of fragmentation, and the latest observation. As one can see, the Afρ parameter is not a constant value at different apertures, like it is expected in a steady-state coma. Thus, Fig. 7 unambiguously suggests that such assumption is not valid in the case of C/2011 J2; for instance, it could imply a pulse-like activity of the nucleus.

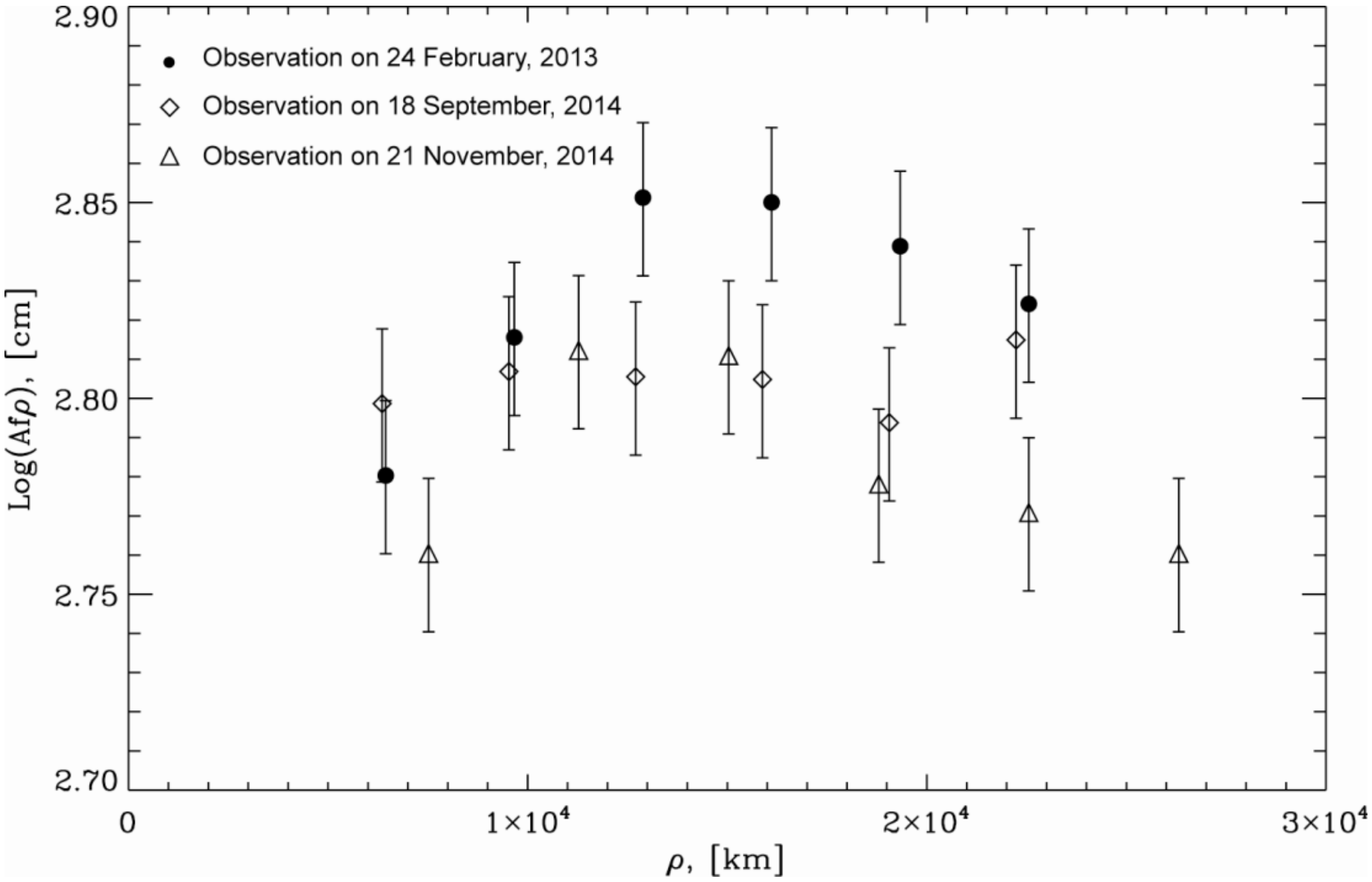

Fig. 7. R-band Afρ trend versus cometocentric distance measured for different date of observations with AZT8 telescope.

### 3.2.3 Dust production rate

We estimate the dust mass loss rate of the comet using the method described in Newburn and Spinrad (1985), Weiler et al. (2003), and taking into account the discussion in Fink and Rubin (2012). We use the obtained Afρ parameters to calculate the dust mass production rate. The relation to estimate the dust productivity is given by

$$Q_M = Q_N (4\pi/3) \cdot \left[ \int_{a_{\min}}^{a_{\max}} \rho_d(a) \cdot a^3 \cdot f(a) da \right], \quad (5)$$

We use the bulk material density of the agglomerated debris particles 0.35–0.83 g cm$^{-3}$ (Zubko et al., 2014) that is consistent with the range 0.3–3 g·cm$^{-3}$, detected in situ in Comet 81P/Wild 2 (Hörz et al., 2006; Flynn, 2008).

Investigations of comets 1P/Halley and 81P/Wild 2 in situ suggest that dust particles obey a power-law size distribution $a^{-n}$, where a is the radius of the equal mass sphere. The power index n is not a constant, it grows with the particle size. In comet 1P/Halley, submicron- and micron-sized dust particles have revealed the power index n = 1.5–3; whereas, in the case of dust particles considerably larger than 1 μm, n = 3.4 (Mazets et al., 1986; 1987). Analysis of data obtained from 81P/Wild 2 reveals that the micron-sized particles obey a power-law size distribution with an index n ≈ 2.9 (Price et al., 2010). However, when large dust particles (≥10 μm) are incorporated into the analysis, it stretches the power index to n = 3.25 – 4.4 (Tuzzolino et al., 2004). For our estimation we use two different functions for size distribution of dust particles: (1) f(a)~$a^{-3.0}$ and (2) f(a)~$\left(1-\frac{a_0}{a}\right)^M \cdot \left(\frac{a_0}{a}\right)^N$, where M=1.8, N=3.6 and $a_0$=5μm (Hanner, 1983). We calculated distribution with lower and upper limits on dust grain radii taken at 0.1 and 1000 μm, respectively.

The dust number rate is given by

$$Q_N = Af\rho \cdot \left[2\pi p(\lambda)\Phi(\alpha)\right]^{-1} \cdot \left[ \int_{a_{\min}}^{a_{\max}} (f(a) \cdot a^2 / v(a)) da \right]^{-1}, \quad (6)$$

where $v(a)$ is the ejection velocity of dust particle. The outflow dust velocity ranged from 1 to 50 $m \cdot s^{-1}$ was based on estimations from numerical modelling of comet dust environment at heliocentric distance being larger than 4 AU (Fulle, 1994; Mazzotta Epifani et al., 2009; Korsun et al., 2010, 2014; Rousselot et al., 2014).

Comet C/2011 J2 was observed over the phase angles 7 – 16.5 degree. For such range, we constrain reflectance of dust particles to the range A = [0.03; 0.07] that is retrieved from modeling of photo-polarimetric observations of comets (Zubko et al. 2014; 2015). These values are generally consistent with the reflectance inferred in other studies (Hanner and Newburn, 1989) and suggested in studies of distant comets (Meech et al., 2009).

The numerical results of the dust mass loss rate are presented in Tab.3. Figure 8 shows results of estimation of dust productivity for the comet C/2011 J2 (LINEAR).

Table 3. Dust mass production rate of the comet C/2011 J2 (LINEAR).

| r, [AU] | $Q_M$, $kg \cdot s^{-1}$ | | | |
|---|---|---|---|---|
| | A=0.03 | | A=0.07 | |
| | A | B | A | B |
| 4.41 | 75.1 | 121.2 | 32.2 | 51.9 |
| 4.21 | 111.5 | 180.0 | 47.8 | 72.1 |
| 4.13 | 186.2 | 300.5 | 79.8 | 128.8 |
| 4.09 | 143.7 | 231.9 | 61.6 | 99.4 |
| 4.00 | 159.6 | 257.6 | 68.4 | 110.4 |
| 3.96 | 91.5 | 147.7 | 39.2 | 63.3 |
| 3.46 | 151.3 | 244.2 | 64.8 | 104.6 |
| 3.51 | 148.8 | 240.2 | 63.8 | 102.9 |
| 3.56 | 183.7 | 296.5 | 78.7 | 127.1 |
| 3.81 | 180.8 | 291.7 | 77.9 | 125.1 |
| 3.86 | 177.3 | 286.2 | 76.1 | 122.6 |
| 4.06 | 123.8 | 199.8 | 53.1 | 85.6 |
| 4.08 | 91.7 | 148.1 | 39.3 | 63.4 |
| 4.10 | 114.8 | 185.2 | 49.2 | 79.4 |
| 4.11 | 94.9 | 153.2 | 40.6 | 65.6 |
| 4.21 | 95.2 | 153.6 | 40.8 | 65.8 |
| 4.22 | 86.7 | 140.0 | 37.1 | 60.0 |
| 4.22 | 82.8 | 133.7 | 35.5 | 57.3 |
| 4.24 | 77.4 | 125.1 | 33.2 | 53.6 |
| 4.25 | 70.6 | 113.9 | 30.2 | 48.8 |
| 4.26 | 68.8 | 111.0 | 29.5 | 47.6 |
| 4.26 | 68.9 | 111.2 | 29.5 | 47.6 |
| 4.30 | 90.2 | 145.6 | 62.4 | 38.7 |
| 4.32 | 93.8 | 151.5 | 40.2 | 64.9 |
| 4.33 | 79.3 | 128.0 | 34.0 | 54.8 |
| 4.34 | 77.2 | 124.6 | 33.1 | 53.4 |
| 4.34 | 83.1 | 134.1 | 35.6 | 57.5 |
| 4.35 | 113.3 | 182.9 | 48.5 | 78.4 |
| 4.35 | 113.1 | 182.7 | 48.3 | 78.3 |
| 4.36 | 103.2 | 166.6 | 44.2 | 71.4 |
| 4.36 | 103.3 | 166.7 | 44.2 | 71.5 |
| 4.37 | 103.2 | 116.5 | 44.2 | 71.3 |
| 4.45 | 90.3 | 145.7 | 38.6 | 62.4 |
| 4.47 | 70.2 | 113.3 | 30.1 | 48.5 |
| 4.55 | 67.4 | 108.7 | 28.8 | 46.6 |
| 4.57 | 87.2 | 140.7 | 37.4 | 60.3 |
| 4.57 | 79.6 | 128.5 | 34.1 | 55.1 |
| 4.58 | 71.5 | 115.4 | 30.6 | 49.4 |
| 4.59 | 67.4 | 108.8 | 28.8 | 46.6 |
| 4.61 | 65.4 | 105.5 | 28.0 | 45.2 |

A- Using dust size distribution function in form $f(a) \sim a^{-3.0}$

B- Using dust size distribution function in form $f(a) \sim \left(1-\frac{a_0}{a}\right)^M \cdot \left(\frac{a_0}{a}\right)^N$

$$M = 1.8, N = 3.6, a_0 = 5\mu m$$

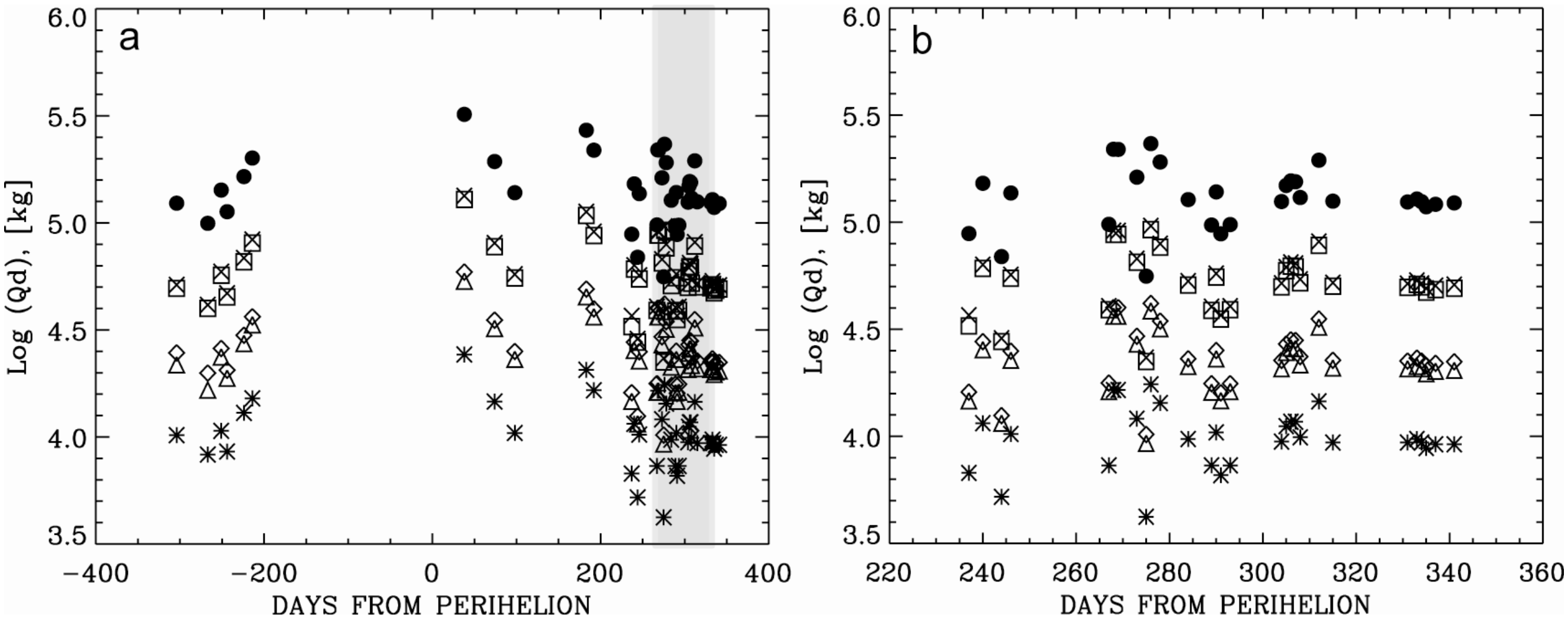

Fig.8 Dependence of the Log of dust productivity $Q_d$ on the time elapsed (a) since perihelion for Comet C/2011 J2 (LINEAR). Asterisk and square triangle mark the $Q_d$ values derived for dust size distribution function in form proposed by Hanner (1983) and for values of the albedo 0.03 and 0.07, respectively. Dots and diamond mark the $Q_d$ values derived for dust size distribution function in form $a^{-3.0}$ and for values of the albedo 0.03 and 0.07, respectively. The grey-shaded area designates the period when the fragmentation was observed. Panel (b) shows in details the variation of the $Q_d$ parameter over the time when fragmentation was apparent.

### 3.3 Optical depth and number density of dust particles in the C/2011 J2 coma

Brightness of comets is governed by a product of reflectance A(α) of their dust and the so-called filling factor f that is a fraction of field of view being overlapped with dust particles. Therefore, the filling factor is defined as a ratio of total geometric cross section of all dust particles within the field of view over the area of the field of view. Evidently, the filling factor corresponds to number density of dust particles. While the product A(α)f can be directly measured, poor knowledge on reflectance of dust particles (for instance, we adapt in this manuscript the range A(5°–15°) = 0.03 – 0.07) places uncertainty in retrieval of the filling factor f.

However, stellar occultation makes it possible an independent estimation of the filling factor. Starlight passing through a cometary coma toward the observer gets attenuated due to interaction with its dust particles and, in general case, gases. Decrement of the starlight flux is dependent on number density of dust particles in coma and their ability to absorb and scatter the incident light in directions other than the forward scattering. The latter capabilities of a dust particle are characterized with the efficiency for extinction $Q_{ext}$ (Bohren & Huffman, 1983). Below, we discuss this characteristic in Fig. 10; whereas, here, one only needs to note a weak dependence of $Q_{ext}$ on material absorption of dust particles that is described with the imaginary part of the complex refractive index m. In other words, the retrieval of number density of dust particles from extinction of starlight is weakly affected by assumption on chemical composition of dust particles, except maybe for the case of optically soft materials, such as water ice. This makes significant difference with the retrieval from apparent magnitude of coma, which is strongly dependent on assumption on reflectance of dust particles. Unfortunately, transit of comet near a bright star is a rare event. Nevertheless, whenever this is possible, it is of high practical interest to compare number density of dust particles in coma inferred with alternative techniques.

Observation of stellar transit near the nucleus was not scheduled as a primary scientific goal in our survey of Comet C/2011 J2 (LINEAR), it became possible incidentally though. On 28 September, 2014 the comet C/2011 J2 (LINEAR) passed near the star USNO-A2 1275-18299027 (it is also known as USNO-B1 star 1345-0506546), when the heliocentric and geocentric distances of the comet were 4.26 and 3.49 AU, respectively. The USNO-B1 magnitude of the star is $16.58^m$in the R filter. The geometry of the event is shown in Fig. 9.

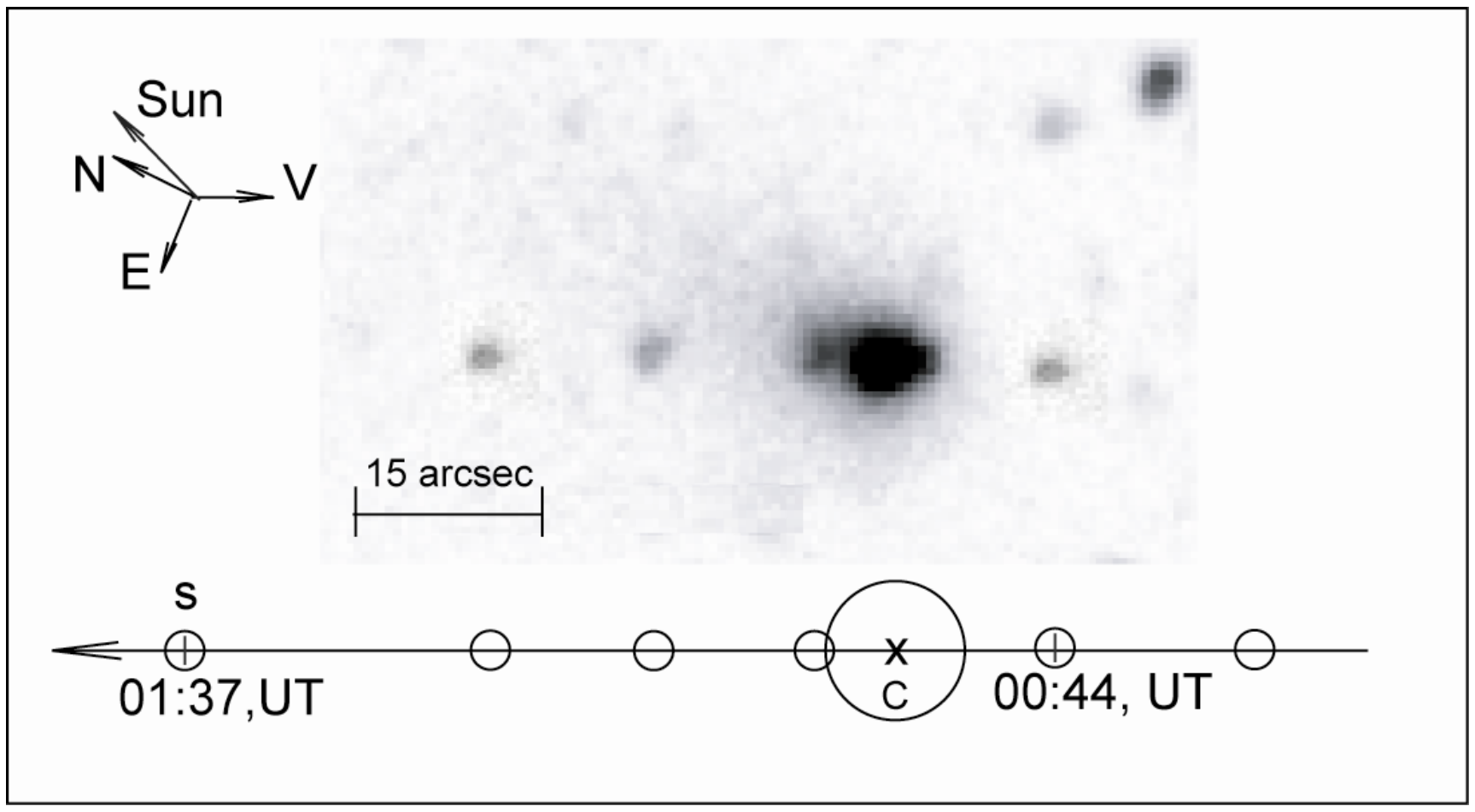

Fig.9. Four CCD frames overlapped together show the star (s) and comet C/2011J2 (LINEAR) (C). The schematic shows the track of the star (during all observations) at 00:44 and 01:37 UT on 28 September, 2014, the scale, and the orientation.

At the moment of occultation, cometary coma is imposed over the star. Therefore, the total observed intensity $I_{sum}$ at this point is equal sum of emission from the star $I_s \cdot e^{-\tau}$ , which weakened by absorption, and emission from comet $I_{com}$

$$I_{sum} = I_s e^{-\tau} + I_{com} , \quad (7)$$

where $I_s$ is intensity of a non-obstructed star. We can obtain the optical depth of the cometary coma from

$$\tau = ln \frac{I_s}{I_{sum} - I_{com}} \tag{8}$$

Comet C/2011 J2 was photometricaly stable, revealing no fast variations of its brightness. Variations of its brightness on transit epoch are constrained to be less than 3.5 per cent. Using the aperture radius ρ= 5062.4 km (2 arcsec), we estimate the optical depth to be $\tau = 0.034 \pm 0.1$. Interestingly, for same aperture, we obtain Afρ=653 cm.

Optical depth τ is a dimensionless characteristic that in the case of optically thin clouds consisting of monodisperse particles and is defined as follows (Mishchenko, 2014):

$$\tau = n \times Q_{ext} \times G. \tag{9}$$

Here, n stands for number of particles in a column with the projected area of 1 $m^2$, $Q_{ext}$ and G – efficiency for extinction [dimensionless] and ensemble-averaged geometric cross section of agglomerated debris particle [in $m^2$].

In Fig. 10 we demonstrate computational results for $Q_{ext}$ in the agglomerated debris particles. These particles have highly irregular agglomerate morphology with packing density of constituent material being consistent with in situ findings in Comet 81P/Wild 2. It is significant that the agglomerated debris particles are capable to reproduce polarization, color, and phase function measured in comets (Zubko et al., 2014; 2015). $Q_{ext}$ in Fig. 10 is shown as a function of the so-called size parameter x (Bohren & Huffman, 1983) that is defined as follows: $x = 2\pi r_{ad}/\lambda$; where $r_{ad}$ – radius of sphere circumscribed around the agglomerated debris particles and λ – wavelength of incident light.

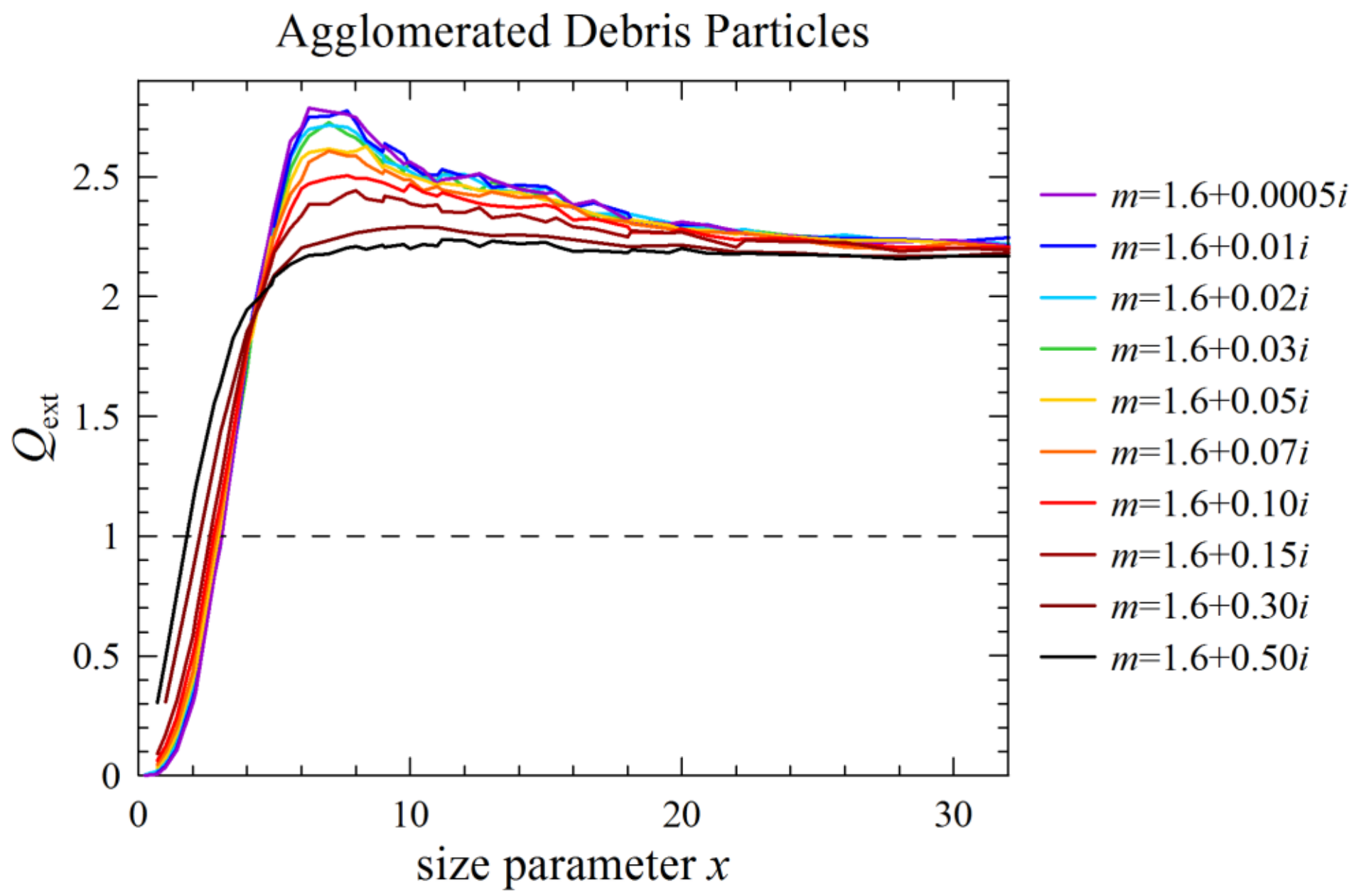

Fig. 10. Efficiency for radiation pressure $Q_{ext}$ versus size parameter x in agglomerated debris particles with different imaginary parts of refractive index m.

As one can see in Fig. 10, $Q_{ext}$ reveals quite similar trends in particles with different levels of material absorption. This implies our estimation of number of dust particles is only weakly dependent upon their chemical composition. However, size parameter x dramatically affects $Q_{ext}$ at $x < 10$; whereas, larger particles are only slightly dependent on x. It is also important to note that the curves in Fig. 10 are quite smoothed in appearance. This behavior differs from highly resonant response in single spherical particles (Bohren & Huffman, 1983).

We compute number density of particles n in assumption of maximum efficiency for extinction $Q_{ext} \approx 2.8$ that corresponds with the bottom limit of the number density n. The maximum value of $Q_{ext}$ occurs at $x = 7$ (see Fig. 10). At wavelength $\lambda = 0.65$ µm (R filter), $x = 7$ is equivalent to $r_{ad} \approx 0.724 \times 10^{-6}$ m. Projected area of the agglomerated debris particles is for a factor of 0.61 smaller than what in the circumscribing sphere, so $G \approx 10^{-12}$ m$^2$. When substituting these values of $Q_{ext}$ and G into eq. (9), we obtain the number density n spanning the range from 0 through $2.5 \times 10^{10}$ particles per 1 m$^2$ of the projected area of coma. This range results from rather large error bars in the measured optical depth $\tau$ =

0.034 ± 0.1. Nevertheless, it is consistent with the number density that can be alternatively inferred from the Afρ parameter.

For instance, during the time of the star transit, Comet C/2011 J2 (LINEAR) has revealed Afρ = 653 cm with the aperture radius ρ = 5062.4 km; this means Af ≈ $1.3 \times 10^{-6}$. In assumption of reflectance of dust particles A = 0.03 – 0.07, the filling factor is equal to f ≈ $(1.843 - 4.300) \times 10^{-5}$. At $r_{ad} \approx 0.724 \times 10^{-6}$ m, this implies the number density n = $(1.833 - 4.278) \times 10^{7}$ particles per 1 $m^2$ of the projected area. This range of the number density n is remarkably lower (for three orders of magnitude) than the upper limit of the range inferred from the optical depth τ. Nevertheless, it is significant that the number density n estimated from the Afρ parameter is consistent with the range of n retrieved from τ.

## 4. Summary and conclusions

We performed R broadband photometry monitoring of Comet C/2011 J2 (LINEAR) observed at the heliocentric distances r = 3.4 – 4.6 AU, with different telescopes, prior and after the perihelion passage. Over the observational period, the comet showed a high level of cometary activity and, also, it experienced a fragmentation. Our principal findings can be summarized as follows.

(1) Analysis of the coma morphology unambiguously reveals fragmentation of Comet C/2011 J2 (LINEAR). We observed the fragment B between September 18, 2014 to November 5, 2014. This goes along with findings by other, independent observers. The endurance of the fragment B is estimated with eq. (1) to be E=29.1 days, which is clearly not consistent with the fact that the fragment B remained observable for some 50 days. Relative velocity of the fragment B is found to be of 0.21″/day (~4.9 m/s) in all our observations.

(2) Upper limit of radius of the comet C/2011 J2 (LINEAR) nucleus is constrained to ≤ 9 km.

(3) The comet revealed a high level of activity for the given range of heliocentric distances that is typical for new comets. The Afρ parameter (Fig.3) and the dust production (Tab. 3) in Comet

C/2011 J2 are comparable to what was previously found in vast majority of dynamically new and long-period comets (Mazzotta Epifani et al. 2009; 2010; Meech et al. 2009; Szabó et al., 2001; 2002; 2008; Korsun et al., 2010; Ivanova et al., 2014). The dust production of the comet differs on epochs before and after the perihelion passage. The activity of the comet was high after the perihelion passage and it was changed during the fragmentation. These results are consistent with findings of the CARA project for this comet.

(4) The stellar near the cometary nucleus made it possible to estimate the optical depth of the coma $\tau$ = 0.034 ± 0.1. In assumption of the particle radius being 0.724 um, we infer the number density of dust particles spanning the range between 0 through $2.5 \times 10^{10}$ particles per 1 $m^2$ of the projected area of coma. Interestingly, the number density n retrieved alternatively from the Afρ parameter appears within this range.

Acknowledgments

This work was supported in part by the Slovak Grant Agency VEGA (grant No. 2/0032/14) and the implementation of the project SAIA. We thank the CARA project members – G. Milani, G. Sostero and E. Bryssink for providing Afρ measurements for the comet. We thank Dr. Neslušan for discussion on calculation of orbital parameters of the comet. We also thank anonymous reviewer for the constructive review.

Highlights

- R band photometry of the comet C/2011 J2 (LINEAR) in the period of its activity in 2013-2014
- The split of the comet detected
- Effective radius of the comet determined
- The total mass of ejected dust estimated
- The stellar occultation of the comet observed; the optical depth of the coma estimated